\documentclass[fleqn]{article}
\usepackage{longtable}
\usepackage{graphicx}

\begin {document}
\begin{flushleft}
{\LARGE

{\bf Radiative rates for E1, E2, M1, and M2 transitions among the 3s$^2$3p$^5$, 3s3p$^6$, and 3s$^2$3p$^4$3d configurations of Cl-like W LVIII} 
} \\

\vspace{1.5 cm}

{\bf {Kanti  M  ~Aggarwal and  Francis   P   ~Keenan}}\\ 

\vspace*{1.0cm}

Astrophysics Research Centre, School of Mathematics and Physics, Queen's University Belfast, Belfast BT7 1NN, Northern Ireland, UK\\ 
\vspace*{0.5 cm} 

e-mail: K.Aggarwal@qub.ac.uk \\

\vspace*{0.20cm}

Received  13 November 2013.  Accepted 16 January 2014 \\

\vspace*{1.5cm}

PACS Ref: 31.25 Jf, 32.70 Cs,  95.30 Ky

\vspace*{1.0 cm}

\hrule

\vspace{0.5 cm}

\end{flushleft}

\clearpage


\begin{abstract}

We report  calculations of energy levels, radiative decay rates, and lifetimes for transitions  among the 3s$^2$3p$^5$, 3s3p$^6$, and 3s$^2$3p$^4$3d configurations of  Cl-like W LVIII. The general-purpose relativistic atomic structure package ({\sc grasp}) has been adopted for our calculations. Comparisons are made with  the most recent results of Mohan et al. [Can. J. Phys. {\bf 92} (2014) xxx]  and discrepancies in  lifetimes are noted, up to  four orders of magnitude in some instances.  Our energy levels are estimated to be accurate to better than 0.5\%, whereas results for radiative rates and lifetimes should be  accurate to  better than 20\%.

\end{abstract}

\clearpage

\section{Introduction}

Atomic data, including  energy levels and  oscillator strengths (or radiative decay rates),  for tungsten (W) ions have been of great interest, both theoretically  \cite{kbf}--\cite{pq} and experimentally  \cite{sbu}--\cite{ll1}, for  several years. This is mainly because tungsten is one of the constituents of tokamak reactor walls. Due to high temperatures of such fusion plasmas, many ionisation stages of W are observable and atomic data are required to assess  radiation losses. The need for atomic data for W ions has become more urgent due to the developing  ITER project. However, there are two requirements from the user community, namely the data should be generated for a significantly large model \cite{dal} and be reliable \cite{fst}. A complete set of data from a large model takes account of cascading effects, while  reliability (based on rigorous tests) provides confidence to the users. 

In a recent paper, Mohan et al.  \cite{mas} have reported results for energy levels, oscillator strengths, radiative rates,  line strengths, and lifetimes for Cl-like W LVIII. For the calculations, they adopted the widely used and readily available {\sc grasp} (general-purpose relativistic atomic structure package) code. This  was originally developed by Grant  et al.  \cite{grasp0} and has undergone several revisions by the names GRASP \cite{grasp}, GRASP2 \cite{grasp92}, and GRASP2K \cite{grasp2k}--\cite{grasp2kk}. The version they  used has been revised by one of its authors (Dr. P. H. Norrington), is known as GRASP0,  and is available at  {\tt http://web.am.qub.ac.uk/DARC/}. It is a fully relativistic code,  based on the $jj$ coupling scheme. Further relativistic corrections arising from the Breit interaction and QED (quantum electrodynamics) effects have also been included. Furthermore, this version provides compatible results with others.

Tungsten is a heavy element (Z = 74) and therefore relativistic effects are very important in the determination of its atomic structure. However, equally important is the inclusion of {\em configuration interaction} (CI) --  see for example, Fournier \cite{kbf} and Aggarwal and Keenan \cite{w40}. In particular, Cl-like ions are very complicated and CI is highly important, as demonstrated by Aggarwal and Keenan for Ti VI \cite{tivi}, Cr VIII \cite{cr8}, Fe X \cite{fex}, and Co XI  \cite{coxi}.  These are comparatively light ions but W LVIII is no exception, and therefore Mohan et al.  \cite{mas} have included CI among 15 configurations (GRASP1), namely 3s$^2$3p$^5$,  3s3p$^6$, 3s$^2$3p$^4$3d, 3s$^2$3p$^4$4$\ell$, 3s3p$^5$3d, 3s$^2$3p$^3$3d$^2$, 3s3p$^4$3d$^2$, 3p$^6$3d,   3s$^2$3p$^2$3d$^3$,  3p$^5$3d$^2$,  3s$^2$3p$^3$4d$^2$, and   3s$^2$3p$^3$4f$^2$. However, inclusion of this limited CI is not sufficient for an accurate determination of energy levels and radiative rates (A- values). Furthermore, they reported energies for only 31 levels of the 3s$^2$3p$^5$, 3s3p$^6$, and 3s$^2$3p$^4$3d  configurations, and A- values for only transitions from the two levels of the ground state, i.e.  (3s$^2$3p$^5$) $^2$P$^o_{3/2,1/2}$. Therefore, there is scope for improvement  and to extend the range of  transitions.
 

\section{Energy levels}

For our calculations we have employed the same  {\sc grasp}  code as adopted by Mohan et al.  \cite{mas}. Additionally, we have used the option of {\em extended average level} (EAL),  in which a weighted (proportional to 2$j$+1) trace of the Hamiltonian matrix is minimised. This produces a compromise set of orbitals  comparable to other options, such as {\em average level} (AL), as noted by Aggarwal  et al.  for several ions of Kr \cite{kr} and Xe \cite{xe}.  However, in our calculations we include much more extensive CI than  Mohan et al.  Gradually increasing the amount of CI, we have performed a series of calculations  with the {\sc grasp} code,  but focus only on our final results which include CI among 38 configurations (GRASP2), listed in Table 1. These  configurations involve the $n \le$ 5 orbitals and generate 3749 levels in total. Their specific energy ranges are  listed in Table 1, and the configurations included by Mohan et al.  are also marked for  reference. Two of the configurations, included by  Mohan et al., namely (3s$^2$3p$^3$) 4d$^2$ and 4f$^2$,  are  not significant, because their levels lie in a very high energy range, i.e. $\ge$ 413 Ryd or $\sim$ 340 Ryd {\em above} the levels considered by these authors.  Hence, their inclusion has little effect on the lower levels. On the other hand, Mohan et al.  excluded some  important configurations, such as   3s3p$^3$3d$^3$,  3s$^2$3p3d$^4$, and 3s3p$^2$3d$^4$. These  configurations together generate 1716 levels and therefore significantly increase the size of a calculation, especially given that the 15 configurations of Mohan et al. generate only 1163 levels. However, the energy range of these three configurations is 92--196 Ryd, i.e. well below that of 3s$^2$3p$^4$4$\ell$ included by Mohan et al. There are additional  configurations  listed in Table 1 whose levels closely interact, intermix, and influence energies of the lower ones.

In Table 2 we compare our calculated energies for 31 levels of the 3s$^2$3p$^5$, 3s3p$^6$, and 3s$^2$3p$^4$3d configurations with those of Mohan et al.  \cite{mas}. Laboratory measurements have been  compiled by  Kramida \cite{ak1} and the NIST (National Institute of Standards and Technology) team \cite{nist}, which are  also included in the table. As stated earlier, we have performed a series of calculations with increasing CI and one of these include the same 15 configurations (GRASP1) as considered by Mohan et al. Our energies for the levels obtained are comparable with their results listed in Table 2, but here we also list the mixing coefficients from the GRASP1 calculations. We note that Mohan et al.  have listed wavefunction compositions  (square of the mixing coefficient) in their Table 1, and hence provide a ready comparison. As a result of a larger CI included in the present work (GRASP2), the energies for a majority of levels are lower than the GRASP1 data by up to 0.07 Ryd, and hence are (slightly) closer to those of NIST. However, there are still discrepancies  with the NIST listings of up to 0.2 Ryd for some levels, such as 13 (3s$^2$3p$^4$($^3$P)3d $^2$D$_{5/2}$).

The NIST listings \cite{nist} for many levels are not based on direct measurements (see levels with a $\star$ in Table 2), but have been derived from interpolation and extrapolation. Similarly, two of their levels, i.e.  3s$^2$3p$^5$  $^2$P$^o_{1/2}$ and 3s$^2$3p$^4$($^1$S)3d  $^2$D$  _{5/2}$ (10 and 11) have the same energy, whereas in the theoretical work the difference between these is  up to  0.1 Ryd, depending on the calculation. Finally, the NIST energy for the 3s3p$^6$ $^2$S$  _{1/2}$   level (14) is anomalous in comparison to all calculations -- also see Table 2 of \cite{mas} and the present Table 3. For this level the NIST energy is {\em higher} by over 1 Ryd and needs a reassessment. Therefore, due to paucity of measured energy levels, the discrepancy noted with the NIST listings should not be considered an indication of the accuracy of our results.

To  assess the accuracy of our energy levels, we have also performed parallel calculations with  the  {\em Flexible Atomic Code} ({\sc fac}) of Gu \cite{fac},  available from the website {\tt http://sprg.ssl.berkeley.edu/$\sim$mfgu/fac/}. This is also a fully relativistic code which provides  results for energy levels and radiative rates comparable to {\sc grasp}, as already shown for several other ions, see for example:  Aggarwal  et al. for Kr \cite{kr} and Xe \cite{xe} ions, and  more recent work on W XL by Aggarwal and Keenan \cite{w40}. With the same amount of CI as included in GRASP1 and GRASP2, the energy levels from {\sc fac} are comparable as listed in Table 2. This has also been shown by Mohan et al.  \cite{mas} in their Table 2, and hence those results are not included in the present Table 2. However, we have performed four  additional calculations with {\sc fac}, with increasing CI.  These are (i) FAC1,  which includes CI among 4580 levels of the 3*7 and  3s$^2$3p$^4$4$\ell$ configurations, (ii) FAC2, which includes a further 1231 levels of 3s$^2$3p$^4$5$\ell$, 3s3p$^5$4$\ell$, 3s3p$^5$5$\ell$, 3p$^6$4$\ell$,    3p$^6$5$\ell$,  and 3s$^2$3p$^3$3d4$\ell$,  (iii) FAC3, with a total of 9160 levels, the additional ones arising from 3s$^2$3p$^4$6$\ell$, 3s3p$^5$6$\ell$, 3p$^6$6$\ell$, 3s$^2$3p$^3$3d5$\ell$, and 3s$^2$3p$^3$3d6$\ell$, and finally (iv) FAC4, with a further 9299 levels (18,459 in total) of 3s$^2$3p$^3$4$\ell^2$,   3s$^2$3p$^3$5$\ell^2$,  3s$^2$3p$^3$4$\ell$5$\ell$,  3p$^5$4$\ell^2$,    and 3p$^5$3d4$\ell$. We discuss these results below.

In Table 3, we compare our GRASP2 energies with the four calculations from {\sc fac} described above. Also listed are the NIST \cite{nist} compilation for a ready reference. All calculations provide  (nearly) the same orderings, and the differences between GRASP2 and all FAC calculations are less than $\sim$ 0.1 Ryd, indicating that the CI included in our GRASP2 calculations is sufficient as far as the levels of Table 3 are concerned. Generally, the energies obtained with {\sc fac} are lower (by up to $\sim$ 0.1 Ryd), as has also been noted by \cite{mas}.  Based on these and earlier comparisons among a variety of calculations with two independent atomic structure codes and with increasing amount of CI, we assess the accuracy of our energy levels to be $\sim$ 0.1 Ryd, or equivalently 0.5\%.

Finally, we note that some of the levels are highly mixed as shown in the present Table 2 and in Table 1 of  Mohan et al.  {\cite{mas}. Therefore, it is not always possible to provide a unique label for each level, but care has been taken to identify the levels as accurately as possible. Particularly difficult to identify are those levels which are highly mixed, such as 2/5/7/11/12/13 shown in Table 2. Therefore, the best one can say about a level is that it has a particular  $J$ value, as listed in Tables 2 and 3, but there can be disagreements about the configuration assigned to it.  For this reason we differ in the designation of a few levels with those of Mohan et al.,  and examples include levels 5 and 7, i.e. (3s$^2$3p$^4$($^3$P)3d) $^2$F$_{7/2}$ and $^4$D$_{7/2}$.  Ralchenko et al. \cite{yuri} have the same designations for these levels, but Mohan et al. have identified these as 3s$^2$3p$^4$($^3$P)3d $^4$F$_{7/2}$ and 3s$^2$3p$^4$($^1$D)3d $^2$F$_{7/2}$. This is a common problem for many ions in which level mixing is very strong, such as Ti VI \cite{tivi}, Cr VIII \cite{cr8}, Fe X \cite{fex}, and Co XI \cite{coxi}.

\section{Radiative rates}

The absorption oscillator strength ($f_{ij}$), a dimensionless quantity,  and radiative rate A$_{ji}$ (in s$^{-1}$) for a transition $i \to j$ are related by the following expression:

\begin{equation}
f_{ij} = \frac{mc}{8{\pi}^2{e^2}}{\lambda^2_{ji}} \frac{{\omega}_j}{{\omega}_i}A_{ji}
  = 1.49 \times 10^{-16} \lambda^2_{ji} \frac{{\omega}_j}{{\omega}_i} A_{ji}
\end{equation}
where $m$ and $e$ are the electron mass and charge, respectively, $c$  the velocity of light,  $\lambda_{ji}$  the transition energy/wavelength in $\rm \AA$, and $\omega_i$ and $\omega_j$  the statistical weights of the lower $i$ and upper $j$ levels, respectively. This relationship between the A- and f- values is the same irrespective of the type of a transition, such as electric dipole (E1), electric quadrupole (E2),  magnetic dipole (M1), and  magnetic quadrupole (M2). In Table 4 we list our A- values for all four types of transitions among the 31 levels of the  3s$^2$3p$^5$, 3s3p$^6$, and 3s$^2$3p$^4$3d configurations,  listed in Table 2. In addition, we  list the f- values and line strengths (S- values), but for E1 transitions alone. For other transitions these can be easily determined though  Eqs. (1--5) of \cite{tivi}.

The A- and f- values have been calculated in both Babushkin and Coulomb gauges, which are  equivalent to the length and velocity forms in the non-relativistic nomenclature. However,  the data in Table 4 are presented  in the length form alone, as these are considered to be comparatively more accurate.  Furthermore, these results correspond to our GRASP2 calculations and the differences (if any) with the corresponding GRASP1 results reported by  Mohan et al.  {\cite{mas} are not significant for most of the E1 transitions. Indeed the only transition for which the two sets of A- values differ substantially is 1--11 (3s$^2$3p$^5$   $^2$P$^o_{3/2}$ -- 3s$^2$3p$^4$($^1$S)3d  $^2$D$  _{5/2}$), for which A (GRASP2) = 1.65$\times$10$^8$ s$^{-1}$ and  A (GRASP1) = 7.57$\times$10$^7$ s$^{-1}$. However, this is a {\em weak} transition with f $\sim$ 10$^{-5}$. More importantly, Mohan et al.  have reported A- values for only 31 E1 and 40 M2 transitions whereas there are 658 possible transitions, among all four types, i.e. 36 E1, 334 E2, 245 M1, and 43 M2, for which results are  listed in Table 4. Finally, there is no discrepancy for the A- values of the 1--2 and 1--3 E1, and 3--7, 5--7, and 5--8 M1 transitions reported by Ralchenko et al. \cite{yuri}. Based on the comparisons between the GRASP1 and GRASP2 calculations, as well as with FAC (see also Table 5 of \cite{mas}), we estimate the accuracy of our A- values and other related parameters to be better than 20\%. This accuracy estimate is further confirmed by the velocity and length ratios of A- values for E1 transitions -- see Table 3 of \cite{mas}.

\section{Lifetimes}

The lifetime $\tau$ of a level $j$ is defined as follows:

\begin{equation}
{\tau}_j = \frac{1}{{\sum_{i}^{}} A_{ji}}.
\end{equation}

In Table 5 we list lifetimes ($\tau$) for all 31 levels from our calculations with the {\sc grasp} code. Results corresponding to our GRASP1 and GRASP2  calculations are listed, and  {\em include} A- values from all types of transitions. Also listed are the lifetimes of Mohan et al.  {\cite{mas}, which correspond to the GRASP1 calculations. As in the present work,  they  have included the contribution of A- values from all four types of transitions. The discrepancies between our  calculated lifetimes and those of  Mohan et al.  are up to four orders of magnitude for about 30\% of  the levels, such as 8, 18, 22, 25, and 30. Although our GRASP2 calculations include a larger CI, the $\tau$ from our GRASP1 are comparable for all levels, except one, namely 3s$^2$3p$^4$($^1$S)3d  $^2$D$  _{5/2}$ (11). This is a direct consequence of the corresponding difference in the A- values of the 1--11 E1  transition, noted in section 3.

To understand the differences in $\tau$ between our calculations and those of Mohan et al.  {\cite{mas}, we also list in  Table 5 the A- values for the {\em dominant} transitions. The only level for which  the E1 transition dominates and the difference in $\tau$  between the two calculations is  four orders of magnitude is 30 (3s$^2$3p$^4$($^3$P)3d  $^4$F$  _{3/2}$). For this level, the A- value for our 10--30 E1 transition  from  GRASP2 is 2.48$\times$10$^{12}$ s$^{-1}$, the same as from GRASP1,  although  Mohan et al.  have not listed an A- value for this transition. Therefore, their reported $\tau$ of 1.10$\times$10$^{-9}$ s is unrealistic. Similarly, for level 18 (3s$^2$3p$^4$($^1$D)3d  $^2$G$  _{7/2}$),  the 5--18 M1 transition dominates with A = 1.60$\times$10$^{8}$ s$^{-1}$ and  $\tau$ = 4.6$\times$10$^{-9}$ s , compared  to $\tau$ = 5.7$\times$10$^{-5}$ s of Mohan et al. Therefore, based on this discussion and the comparisons shown in Table 5, we can confidently state that some of the lifetimes reported by Mohan et al. are  overestimated. Finally,   there are no measurements  available with which to compare our results, but we hope our reported  lifetimes will be helpful to experimentalists. 

\section{Conclusions}

In this work, energy levels, radiative rates, oscillator strengths, and lifetimes have been calculated with the {\sc grasp} code for a large number of levels/transitions of W LVIII, but results have been presented only among the 31 levels of the  3s$^2$3p$^5$, 3s3p$^6$, and 3s$^2$3p$^4$3d configurations, as similar data  have recently been reported by Mohan et al.  {\cite{mas}. These authors included limited CI in the determination of atomic structure and, more importantly,  reported A- values for E1 and M2  transitions only from the levels of the ground state. This limited data is not helpful for modelling of plasmas, as demonstrated in Fig. 2 and Table 6 of Del Zanna et al. \cite{del} for a similar Cl-like ion Fe X. To be specific level populations deduced using a limited data may be different by up to a factor of five. Therefore, in our work A- values have been listed for all possible 658 E1, E2, M1, and M2 transitions, which should be useful for plasma modelling.

As a result of the increased CI included in our work, the accuracy of the energy levels has improved by up to 0.07 Ryd, and brings the theoretical results closer to measurements, although limited to a few levels alone. The accuracy of our data is assessed to be better than 0.5\% (within 0.1 Ryd),  based on a series of calculations performed not only with the {\sc grasp} code, but also with {\sc fac}. Similarly, the accuracy of our radiative rates is assessed to be better than 20\% for a majority of transitions.

Although no significant discrepancies are observed for the A- values reported by Mohan et al.  {\cite{mas} for a limited number of transitions, their calculated lifetimes are in error for several levels, by up to four orders of magnitude. Finally, similar  results for a larger number of levels and their corresponding transitions  will be reported in a later paper.

\section*{Acknowledgment}
 KMA  is thankful to  AWE Aldermaston for financial support.     



\newpage
\clearpage

\begin{flushleft}
Table 1.Configurations/Levels of W LVIII and their energy ranges (Ryd). 
\end{flushleft}
\begin{tabular}{rllcrrrrrrrrr} \hline
\\
Index  & Configuration      & No. of Levels  & Energy Range   &  GRASP1 & GRASP2    \\
\\ \hline
  1  &  3s$^2$3p$^5$	      &     2$^o$    & 0-26	  &  Y &  Y  &  &   &   &    \\
  2  &  3s$^2$3p$^4$3d        &     28       & 17-75	  &  Y &  Y  &  &   &   &    \\
  3  &  3s3p$^6$	      &     1	     & 40	  &  Y &  Y  &  &   &   &    \\
  4  &  3s$^2$3p$^3$3d$^2$    &   141$^o$    & 35-103	  &  Y &  Y  &  &   &   &    \\
  5  &  3s3p$^5$3d	      &    23$^o$    & 55-91	  &  Y &  Y  &  &   &   &    \\
  6  &  3s$^2$3p$^2$3d$^3$    &    261       & 55-130	  &  Y &  Y  &  &   &   &    \\
  7  &  3s3p$^4$3d$^2$        &    211	     & 73-143	  &  Y &  Y  &  &   &   &    \\
  8  &  3s3p$^3$3d$^3$        &    678$^o$   & 92-169	  &    &  Y  &  &   &   &    \\
  9  &  3p$^6$3d	      &    2	     & 100-106    &  Y &  Y  &  &   &   &    \\
 10  &  3s$^2$3p3d$^4$        &    180$^o$   & 102-155    &    &  Y  &  &   &   &    \\
 11  &  3s3p$^2$3d$^4$        &    858       & 113-196    &    &  Y  &  &   &   &    \\
 12  &  3p$^5$3d$^2$	      &    45$^o$    & 116-157    &  Y &  Y  &  &   &   &    \\
 13  &  3s$^2$3p$^4$4s        &     8	     & 182-209    &  Y &  Y  &  &   &   &    \\
 14  &  3s$^2$3p$^4$4p        &     21$^o$   & 188-251    &  Y &  Y  &  &   &   &    \\
 15  &  3s$^2$3p$^4$4d        &     28       & 205-261    &  Y &  Y  &  &   &   &    \\
 16  &  3s$^2$3p$^4$4f        &    30$^o$    & 214-268    &  Y &  Y  &  &   &   &    \\
 17  &  3s$^2$3p$^3$3d4s      &    72$^o$    & 199-262    &    &  Y  &  &   &   &    \\
 18  &  3s$^2$3p$^3$3d4p      &    203       & 205-278    &    &  Y  &  &   &   &    \\
 19  &  3s$^2$3p$^3$3d4d      &    302$^o$   & 222-288    &    &  Y  &  &   &   &    \\
 20  &  3s$^2$3p$^3$3d4f      &    363       & 231-295    &    &  Y  &  &   &   &    \\
 21  &  3s3p$^5$4s	      &    7$^o$     & 221-250    &    &  Y  &  &   &   &    \\
 22  &  3s3p$^5$4p	      &    18	     & 227-265    &    &  Y  &  &   &   &    \\
 23  &  3s3p$^5$4d	      &    23$^o$    & 244-275    &    &  Y  &  &   &   &    \\
 24  &  3s3p$^5$4f	      &    24	     & 253-283    &    &  Y  &  &   &   &    \\
 25  &  3s$^2$3p$^4$5s        &    8	     & 276-329    &    &  Y  &  &   &   &    \\
 26  &  3s$^2$3p$^4$5p        &    21$^o$    & 279-337    &    &  Y  &  &   &   &    \\
 27  &  3s$^2$3p$^4$5d        &    28	     & 288-342    &    &  Y  &  &   &   &    \\
 28  &  3s$^2$3p$^4$5f        &    30$^o$    & 292-346    &    &  Y  &  &   &   &    \\
 29  &  3s$^2$3p$^4$5g        &    30	     & 295-348    &    &  Y  &  &   &   &    \\
 30  &  3p$^6$4s	      &     1	     & 267	  &    &  Y  &  &   &   &    \\
 31  &  3p$^6$4p	      &    2$^o$     & 271-283    &    &  Y  &  &   &   &    \\
 32  &  3p$^6$4d	      &    2	     & 290-293    &    &  Y  &  &   &   &    \\
 33  &  3p$^6$4f	      &    2$^o$     & 298-300    &    &  Y  &  &   &   &    \\
 34  &  3s3p$^5$5s	      &    7	     & 316-345    &    &  Y  &  &   &   &    \\
 35  &  3s3p$^5$5p	      &    18	     & 320-353    &    &  Y  &  &   &   &    \\
 36  &  3s3p$^5$5d	      &    23$^o$    & 328-358    &    &  Y  &  &   &   &    \\
 37  &  3s3p$^5$5f	      &    24	     & 332-361    &    &  Y  &  &   &   &    \\
 38  &  3s3p$^5$5g	      &    24$^o$    & 334-363    &    &  Y  &  &   &   &    \\
 39  &  3s$^2$3p$^3$4d$^2$    &    141$^o$   & 413-472    &  Y &     &  &   &   &    \\
 40  &  3s$^2$3p$^3$4f$^2$    &    221$^o$   & 430-487    &  Y &     &  &   &   &    \\
 \\ \hline  											      
\end{tabular}   								   					       
			      							   					       
\begin{flushleft}													       
{\small
GRASP1:  earlier calculations of Mohan et al. \cite{mas}  from the {\sc grasp} code with 1163 levels \\ 
GRASP2: present calculations from the {\sc grasp} code with 3749 levels\\
Y: configuration included under a calculation \\ 										       
															       
}															       
\end{flushleft} 

\newpage
\clearpage

\begin{flushleft}
Table 2. Energies (Ryd) for some levels of W LVIII and their mixing coefficients (MC). 
\end{flushleft}
\begin{tabular}{rlllrrrrrrrrr} \hline
 & & & & & & & &  \\
Index  & Configuration             & Level              & NIST   &  GRASP1  & GRASP2     & MC    \\
& & & & & & & &   \\ \hline
& & & & & & & &   \\
  1  &  3s$^2$3p$^5$               &  $^2$P$^o_{3/2}$	&  00.000 &  0.0000  &   0.0000   &  1( 0.995)			       \\ 
  2  &  3s$^2$3p$^4$($^3$P)3d 	   &  $^4$D$  _{3/2}$	& 17.344 & 17.4534  &  17.4004   &  2(-0.595)+16( 0.455)+12( 0.442)    \\
  3  &  3s$^2$3p$^4$($^3$P)3d      &  $^4$D$  _{5/2}$	& 17.794 & 17.9141  &  17.8572   &  3( 0.614)+19( 0.447)	       \\
  4  &  3s$^2$3p$^4$($^3$P)3d      &  $^4$P$  _{1/2}$	& 17.870$^\star$ & 17.9752  &  17.9196   &  4(-0.708)+29( 0.461)	       \\
  5  &  3s$^2$3p$^4$($^3$P)3d      &  $^2$F$  _{7/2}$	& 18.051 & 18.1905  &  18.1269   &  5( 0.566)+22( 0.539)+18( 0.534)    \\
  6  &  3s$^2$3p$^4$($^1$S)3d      &  $^2$D$  _{3/2}$	& 19.460$^\star$ & 19.6176  &  19.5587   &  6( 0.758)+30( 0.436)	       \\
  7  &  3s$^2$3p$^4$($^3$P)3d      &  $^4$D$  _{7/2}$	& 23.280 & 23.3713  &  23.3138   &  7( 0.652)+27( 0.502)+5( 0.471)     \\
  8  &  3s$^2$3p$^4$($^3$P)3d      &  $^4$F$  _{9/2}$	& 23.439 & 23.5406  &  23.4800   &  8( 0.824)+25( 0.562)	       \\
  9  &  3s$^2$3p$^4$($^3$P)3d 	   &  $^2$P$  _{1/2}$	& 24.080$^\star$ & 24.0296  &  23.9747   &  9(-0.672)+13( 0.472)	       \\
 10  &  3s$^2$3p$^5$  		   &  $^2$P$^o_{1/2}$	& 25.500$^\star$ & 25.5348  &  25.5359   & 10( 0.993)			       \\
 11  &  3s$^2$3p$^4$($^1$S)3d      &  $^2$D$  _{5/2}$	& 25.500$^\star$ & 25.6854  &  25.6227   & 11( 0.681)+31( 0.465)+24(-0.403)    \\
 12  &  3s$^2$3p$^4$($^3$P)3d      &  $^4$P$  _{3/2}$	& 26.130$^\star$ & 26.1767  &  26.1047   & 12(-0.468)+23( 0.453)+28(-0.444)    \\
 13  &  3s$^2$3p$^4$($^3$P)3d      &  $^2$D$  _{5/2}$	& 26.600$^\star$ & 26.4839  &  26.4171   & 13( 0.518)+3(-0.411)+11(-0.406)     \\
 14  &  3s3p$^6$ 		   &  $^2$S$  _{1/2}$	& 41.200$^\star$ & 39.8947  &  39.8704   & 14( 0.855)			       \\
 15  &  3s$^2$3p$^4$($^3$P)3d      &  $^4$D$  _{1/2}$	& 42.200$^\star$ & 42.1254  &  42.0831   & 15(-0.886)			       \\
 16  &  3s$^2$3p$^4$($^1$D)3d      &  $^2$D$  _{3/2}$	& 42.900$^\star$ & 42.9204  &  42.8705   & 30(-0.571)+2( 0.566)+16( 0.311)     \\
 17  &  3s$^2$3p$^4$($^3$P)3d      &  $^4$F$  _{5/2}$	& 43.500$^\star$ & 43.5241  &  43.4668   & 17( 0.704)+19( 0.460)	       \\
 18  &  3s$^2$3p$^4$($^1$D)3d      &  $^2$G$  _{7/2}$	& 43.670$^\star$ & 43.6284  &  43.5691   & 18(-0.770)+5( 0.413) 	       \\
 19  &  3s$^2$3p$^4$($^1$D)3d      &  $^2$F$  _{5/2}$	& 45.780 & 45.9809  &  45.9158   & 13( 0.579)+19(-0.530)	       \\
 20  &  3s$^2$3p$^4$($^1$D)3d      &  $^2$P$  _{3/2}$	& 45.990 & 46.1664  &  46.0980   & 20( 0.519)+16(-0.495)+12( 0.474)    \\
 21  &  3s$^2$3p$^4$($^1$D)3d      &  $^2$S$  _{1/2}$	& 46.450 & 46.6876  &  46.6326   & 21( 0.517)+29(-0.508)+4(-0.472)     \\
 22  &  3s$^2$3p$^4$($^3$P)3d      &  $^4$F$  _{7/2}$	&	 & 48.3574  &  48.3091   &  7( 0.663)+22( 0.620)	       \\
 23  &  3s$^2$3p$^4$($^3$P)3d      &  $^2$P$  _{3/2}$	& 	 & 49.0564  &  49.0046   & 23( 0.536)+12(-0.448)+20(-0.446)    \\
 24  &  3s$^2$3p$^4$($^3$P)3d      &  $^2$F$  _{5/2}$	& 	 & 49.3626  &  49.3074   & 24( 0.657)+31( 0.473)	       \\
 25  &  3s$^2$3p$^4$($^1$D)3d      &  $^2$G$  _{9/2}$	& 	 & 49.3420  &  49.2842   & 25(-0.822)+8( 0.561) 	       \\
 26  &  3s$^2$3p$^4$($^1$D)3d      &  $^2$D$  _{5/2}$	&	 & 49.8777  &  49.8205   & 26(-0.653)+31( 0.520)	       \\
 27  &  3s$^2$3p$^4$($^1$D)3d      &  $^2$F$  _{7/2}$	&	 & 50.5070  &  50.4437   & 27(-0.747)+5( 0.436) 	       \\
 28  &  3s$^2$3p$^4$($^3$P)3d      &  $^2$D$  _{3/2}$	&	 & 52.2559  &  52.1908   & 28(-0.666)+16(-0.439)	       \\
 29  &  3s$^2$3p$^4$($^1$D)3d      &  $^2$P$  _{1/2}$	&	 & 52.9774  &  52.9102   &  9(-0.571)+29(-0.557)+13(-0.547)    \\
 30  &  3s$^2$3p$^4$($^3$P)3d      &  $^4$F$  _{3/2}$	&	 & 71.0520  &  70.9979   &  6( 0.585)+30(-0.504)+28(-0.405)    \\
 31  &  3s$^2$3p$^4$($^3$P)3d      &  $^4$P$  _{5/2}$	&	 & 75.4505  &  75.4010   & 11( 0.589)+24( 0.419)+31(-0.362)    \\
& & & & & & & & \\ \hline            								                	 
\end{tabular}   								   					       
			      							   					       
\begin{flushleft}													       
{\small
NIST: \cite{nist},  levels with a $\star$ are determined by interpolation/extrapolation\\  
GRASP1: earlier calculations of Mohan et al. \cite{mas}  from the {\sc grasp} code with 1163 levels \\ 
GRASP2: present calculations from the {\sc grasp} code with 3749 levels\\
MC: Mixing coefficients corresponding to GRASP1 calculations \\															       
}															       
\end{flushleft} 

\newpage
\clearpage
 
\begin{flushleft}
Table 3. Comparison of energies (Ryd) from the {\sc grasp} and {\sc fac} codes for some levels of W LVIII. 
\end{flushleft}
\begin{tabular}{rlllrrrrrrrrr} \hline
 & & & & & & & &  \\
Index  & Configuration             & Level              & NIST   & GRASP2    &   FAC1     &   FAC2     &    FAC3     &   FAC4      \\
& & & & & & & &   \\ \hline
& & & & & & & &   \\
  1  &  3s$^2$3p$^5$               &  $^2$P$^o_{3/2}$	&  00.000&   0.0000  &   0.00000  &   0.00000  &    0.00000  &   0.00000   \\ 
  2  &  3s$^2$3p$^4$($^3$P)3d 	   &  $^4$D$  _{3/2}$	& 17.344 &  17.4004  &  17.32500  &  17.30007  &   17.29148  &  17.31476   \\
  3  &  3s$^2$3p$^4$($^3$P)3d      &  $^4$D$  _{5/2}$	& 17.794 &  17.8572  &  17.78425  &  17.75610  &   17.74686  &  17.76984   \\
  4  &  3s$^2$3p$^4$($^3$P)3d      &  $^4$P$  _{1/2}$	& 17.870 &  17.9196  &  17.84417  &  17.81809  &   17.80805  &  17.83068   \\
  5  &  3s$^2$3p$^4$($^3$P)3d      &  $^2$F$  _{7/2}$	& 18.051 &  18.1269  &  18.05758  &  18.02481  &   18.01469  &  18.03738   \\
  6  &  3s$^2$3p$^4$($^1$S)3d      &  $^2$D$  _{3/2}$	& 19.460 &  19.5587  &  19.48303  &  19.46102  &   19.45262  &  19.47488   \\
  7  &  3s$^2$3p$^4$($^3$P)3d      &  $^4$D$  _{7/2}$	& 23.280 &  23.3138  &  23.25604  &  23.22210  &   23.21249  &  23.23596   \\
  8  &  3s$^2$3p$^4$($^3$P)3d      &  $^4$F$  _{9/2}$	& 23.439 &  23.4800  &  23.42512  &  23.38844  &   23.37934  &  23.40281   \\
  9  &  3s$^2$3p$^4$($^3$P)3d 	   &  $^2$P$  _{1/2}$	& 24.080 &  23.9747  &  23.91035  &  23.88147  &   23.87052  &  23.89290   \\
 10  &  3s$^2$3p$^5$  		   &  $^2$P$^o_{1/2}$	& 25.500 &  25.5359  &  25.56424  &  25.56100  &   25.56023  &  25.56029   \\
 11  &  3s$^2$3p$^4$($^1$S)3d      &  $^2$D$  _{5/2}$	& 25.500 &  25.6227  &  25.56139  &  25.53224  &   25.52054  &  25.54198   \\
 12  &  3s$^2$3p$^4$($^3$P)3d      &  $^4$P$  _{3/2}$	& 26.130 &  26.1047  &  26.04906  &  26.01088  &   25.99531  &  26.01665   \\
 13  &  3s$^2$3p$^4$($^3$P)3d      &  $^2$D$  _{5/2}$	& 26.600 &  26.4171  &  26.35345  &  26.32222  &   26.31125  &  26.33340   \\
 14  &  3s3p$^6$ 		   &  $^2$S$  _{1/2}$	& 41.200 &  39.8704  &  39.80383  &  39.79163  &   39.79341  &  39.80298   \\
 15  &  3s$^2$3p$^4$($^3$P)3d      &  $^4$D$  _{1/2}$	& 42.200 &  42.0831  &  42.04051  &  42.01204  &   42.00635  &  42.03020   \\
 16  &  3s$^2$3p$^4$($^1$D)3d      &  $^2$D$  _{3/2}$	& 42.900 &  42.8705  &  42.83200  &  42.79886  &   42.79085  &  42.81449   \\
 17  &  3s$^2$3p$^4$($^3$P)3d      &  $^4$F$  _{5/2}$	& 43.500 &  43.4668  &  43.43225  &  43.39470  &   43.38525  &  43.40843   \\
 18  &  3s$^2$3p$^4$($^1$D)3d      &  $^2$G$  _{7/2}$	& 43.670 &  43.5691  &  43.53531  &  43.49602  &   43.48692  &  43.51048   \\
 19  &  3s$^2$3p$^4$($^1$D)3d      &  $^2$F$  _{5/2}$	& 45.780 &  45.9158  &  45.88080  &  45.84312  &   45.83029  &  45.85189   \\
 20  &  3s$^2$3p$^4$($^1$D)3d      &  $^2$P$  _{3/2}$	& 45.990 &  46.0980  &  46.06912  &  46.02731  &   46.01176  &  46.03323   \\
 21  &  3s$^2$3p$^4$($^1$D)3d      &  $^2$S$  _{1/2}$	& 46.450 &  46.6326  &  46.57652  &  46.55101  &   46.54240  &  46.56093   \\
 22  &  3s$^2$3p$^4$($^3$P)3d      &  $^4$F$  _{7/2}$	&	 &  48.3091  &  48.28499  &  48.24739  &   48.23930  &  48.26340   \\
 23  &  3s$^2$3p$^4$($^3$P)3d      &  $^2$P$  _{3/2}$	& 	 &  49.0046  &  48.97824  &  48.94040  &   48.92859  &  48.95174   \\
 24  &  3s$^2$3p$^4$($^3$P)3d      &  $^2$F$  _{5/2}$	& 	 &  49.3074  &  49.28477  &  49.24413  &   49.23220  &  49.25484   \\
 25  &  3s$^2$3p$^4$($^1$D)3d      &  $^2$G$  _{9/2}$	& 	 &  49.2842  &  49.26511  &  49.22123  &   49.21161  &  49.23535   \\
 26  &  3s$^2$3p$^4$($^1$D)3d      &  $^2$D$  _{5/2}$	&	 &  49.8205  &  49.79443  &  49.75444  &   49.74101  &  49.76337   \\
 27  &  3s$^2$3p$^4$($^1$D)3d      &  $^2$F$  _{7/2}$	&	 &  50.4437  &  50.42117  &  50.37700  &   50.36281  &  50.38456   \\
 28  &  3s$^2$3p$^4$($^3$P)3d      &  $^2$D$  _{3/2}$	&	 &  52.1908  &  52.16259  &  52.12437  &   52.11111  &  52.13271   \\
 29  &  3s$^2$3p$^4$($^1$D)3d      &  $^2$P$  _{1/2}$	&	 &  52.9102  &  42.83200  &  42.79886  &   42.79085  &  42.81449   \\
 30  &  3s$^2$3p$^4$($^3$P)3d      &  $^4$F$  _{3/2}$	&	 &  70.9979  &  70.99515  &  70.95580  &   70.94608  &  70.96864   \\
 31  &  3s$^2$3p$^4$($^3$P)3d      &  $^4$P$  _{5/2}$	&	 &  75.4010  &  75.41228  &  75.37291  &   75.36308  &  75.38599   \\
& & & & & & & & \\ \hline            								                	 
\end{tabular}   								   					       
			      							   					       
\begin{flushleft}													       
{\small
NIST: \cite{nist} \\  
GRASP2: present calculations from the {\sc grasp} code with 3749 levels\\
FAC1: 	present calculations from the {\sc fac} code with 4580 levels\\	
FAC2: 	present calculations from the {\sc fac} code with 5821 levels\\	
FAC3: 	present calculations from the {\sc fac} code with 9160 levels\\	
FAC4: 	present calculations from the {\sc fac} code with 18,459 levels\\														       
}															       
\end{flushleft} 

\newpage
\clearpage
													       
\setcounter{table}{3}                                                                                                                                           
\begin{table*}                                                                                                                                                  
\caption{Transition wavelengths in vacuum ($\lambda_{ij}$ in $\rm \AA$), radiative rates (A$_{ji}$ in s$^{-1}$), oscillator strengths (f$_{ij}$, dimensionless), and line      
strengths (S, in atomic units), in Babushkin gauge, for electric dipole (E1), and A$_{ji}$ for E2, M1,  and M2 transitions in W LVIII. ($a{\pm}b \equiv a{\times}$10$^{{\pm}b}$).}
\begin{tabular}{rrrrrrrrrr}                                                                                                                                                    
\hline 
\hline                                                                                                                                                                              
$i$ & $j$ & $\lambda_{ij}$ & A$^{{\rm E1}}_{ji}$  & f$^{{\rm E1}}_{ij}$ & S$^{{\rm E1}}$ & A$^{{\rm E2}}_{ji}$  & A$^{{\rm M1}}_{ji}$ & A$^{{\rm M2}}_{ji}$  \\                     
\hline                                                                                                                                                                        
    1 &    2 &  5.237$+$01 &  2.646$+$09 &  1.088$-$03 &  7.503$-$04 &  0.000$+$00 &  0.000$+$00 &  3.430$+$00 \\                           
    1 &    3 &  5.103$+$01 &  4.156$+$09 &  2.434$-$03 &  1.636$-$03 &  0.000$+$00 &  0.000$+$00 &  1.466$+$01 \\                           
    1 &    4 &  5.085$+$01 &  2.011$+$10 &  3.899$-$03 &  2.611$-$03 &  0.000$+$00 &  0.000$+$00 &  3.639$+$01 \\                           
    1 &    5 &  5.027$+$01 &  0.000$+$00 &  0.000$+$00 &  0.000$+$00 &  0.000$+$00 &  0.000$+$00 &  3.598$+$01 \\                           
    1 &    6 &  4.659$+$01 &  1.836$+$09 &  5.975$-$04 &  3.666$-$04 &  0.000$+$00 &  0.000$+$00 &  1.471$+$02 \\                           
    1 &    7 &  3.909$+$01 &  0.000$+$00 &  0.000$+$00 &  0.000$+$00 &  0.000$+$00 &  0.000$+$00 &  3.608$+$04 \\                           
    1 &    9 &  3.801$+$01 &  8.374$+$10 &  9.069$-$03 &  4.539$-$03 &  0.000$+$00 &  0.000$+$00 &  9.836$+$04 \\
\hline                                                                                                                                                                              
\end{tabular}                                                                                                                                                                       
\end{table*}                                                                                                                                                                        
\newpage                                                                                                                                                                            
\clearpage                                                                                                                                                                          
\setcounter{table}{3}                                                                                                                                                               
\begin{table*}                                                                                                                                                                      
\caption{Transition wavelengths in vacuum ($\lambda_{ij}$ in $\rm \AA$), radiative rates (A$_{ji}$ in s$^{-1}$), oscillator strengths (f$_{ij}$, dimensionless), and line           
strengths (S, in atomic units), in Babushkin gauge, for electric dipole (E1), and A$_{ji}$ for E2, M1, and M2 transitions in W LVIII. ($a{\pm}b \equiv a{\times}$10$^{{\pm}b}$).}   
\begin{tabular}{rrrrrrrrrr}                                                                                                                                                         
\hline                                                                                                                                                                              
\hline                                                                                                                                                                              
$i$ & $j$ & $\lambda_{ij}$ & A$^{{\rm E1}}_{ji}$  & f$^{{\rm E1}}_{ij}$ & S$^{{\rm E1}}$ & A$^{{\rm E2}}_{ji}$  & A$^{{\rm M1}}_{ji}$ & A$^{{\rm M2}}_{ji}$  \\                     
\hline                                                                                                                                                                              
    2 &   16 &  3.578$+$01 &  0.000$+$00 &  0.000$+$00 &  0.000$+$00 &  1.412$+$06 &  2.033$+$06 &  0.000$+$00 \\                           
    2 &   17 &  3.496$+$01 &  0.000$+$00 &  0.000$+$00 &  0.000$+$00 &  2.042$+$06 &  8.734$+$04 &  0.000$+$00 \\                           
    2 &   18 &  3.482$+$01 &  0.000$+$00 &  0.000$+$00 &  0.000$+$00 &  1.296$+$06 &  0.000$+$00 &  0.000$+$00 \\                           
    2 &   19 &  3.196$+$01 &  0.000$+$00 &  0.000$+$00 &  0.000$+$00 &  2.584$+$05 &  7.214$+$07 &  0.000$+$00 \\                           
    2 &   20 &  3.175$+$01 &  0.000$+$00 &  0.000$+$00 &  0.000$+$00 &  3.468$+$04 &  1.157$+$08 &  0.000$+$00 \\                           
    2 &   21 &  3.117$+$01 &  0.000$+$00 &  0.000$+$00 &  0.000$+$00 &  6.462$+$06 &  2.183$+$07 &  0.000$+$00 \\                           
    2 &   22 &  2.948$+$01 &  0.000$+$00 &  0.000$+$00 &  0.000$+$00 &  1.306$+$03 &  0.000$+$00 &  0.000$+$00 \\
\hline                                                                                                                                                                              
\end{tabular}                                                                                                                                                                       
\end{table*}                                                                                                                                                                        
\newpage                                                                                                                                                                            
\clearpage                                                                                                                                                                          
\setcounter{table}{3}                                                                                                                                                               
\begin{table*}                                                                                                                                                                      
\caption{Transition wavelengths in vacuum ($\lambda_{ij}$ in $\rm \AA$), radiative rates (A$_{ji}$ in s$^{-1}$), oscillator strengths (f$_{ij}$, dimensionless), and line           
strengths (S, in atomic units), in Babushkin gauge, for electric dipole (E1), and A$_{ji}$ for E2, M1, and M2 transitions in W LVIII. ($a{\pm}b \equiv a{\times}$10$^{{\pm}b}$).}   
\begin{tabular}{rrrrrrrrrr}                                                                                                                                                         
\hline                                                                                                                                                                              
\hline                                                                                                                                                                              
$i$ & $j$ & $\lambda_{ij}$ & A$^{{\rm E1}}_{ji}$  & f$^{{\rm E1}}_{ij}$ & S$^{{\rm E1}}$ & A$^{{\rm E2}}_{ji}$  & A$^{{\rm M1}}_{ji}$ & A$^{{\rm M2}}_{ji}$  \\                     
\hline                                                                                                                                                                              
    3 &   29 &  2.600$+$01 &  0.000$+$00 &  0.000$+$00 &  0.000$+$00 &  3.524$+$04 &  0.000$+$00 &  0.000$+$00 \\                           
    3 &   30 &  1.715$+$01 &  0.000$+$00 &  0.000$+$00 &  0.000$+$00 &  1.518$+$04 &  4.967$+$03 &  0.000$+$00 \\                           
    3 &   31 &  1.584$+$01 &  0.000$+$00 &  0.000$+$00 &  0.000$+$00 &  1.951$+$04 &  4.917$+$00 &  0.000$+$00 \\                           
    4 &    6 &  5.560$+$02 &  0.000$+$00 &  0.000$+$00 &  0.000$+$00 &  3.253$+$00 &  8.760$+$03 &  0.000$+$00 \\                           
    4 &    9 &  1.505$+$02 &  0.000$+$00 &  0.000$+$00 &  0.000$+$00 &  0.000$+$00 &  3.396$+$06 &  0.000$+$00 \\                           
    4 &   10 &  1.197$+$02 &  5.964$+$07 &  1.280$-$04 &  1.008$-$04 &  0.000$+$00 &  0.000$+$00 &  0.000$+$00 \\                           
    4 &   11 &  1.183$+$02 &  0.000$+$00 &  0.000$+$00 &  0.000$+$00 &  5.807$+$01 &  0.000$+$00 &  0.000$+$00 \\
\hline                                                                                                                                                                              
\end{tabular}                                                                                                                                                                       
\end{table*}                                                                                                                                                                        
\newpage                                                                                                                                                                            
\clearpage                                                                                                                                                                          
\setcounter{table}{3}                                                                                                                                                               
\begin{table*}                                                                                                                                                                      
\caption{Transition wavelengths in vacuum ($\lambda_{ij}$ in $\rm \AA$), radiative rates (A$_{ji}$ in s$^{-1}$), oscillator strengths (f$_{ij}$, dimensionless), and line           
strengths (S, in atomic units), in Babushkin gauge, for electric dipole (E1), and A$_{ji}$ for E2, M1, and M2 transitions in W LVIII. ($a{\pm}b \equiv a{\times}$10$^{{\pm}b}$).}   
\begin{tabular}{rrrrrrrrrr}                                                                                                                                                         
\hline                                                                                                                                                                              
\hline                                                                                                                                                                              
$i$ & $j$ & $\lambda_{ij}$ & A$^{{\rm E1}}_{ji}$  & f$^{{\rm E1}}_{ij}$ & S$^{{\rm E1}}$ & A$^{{\rm E2}}_{ji}$  & A$^{{\rm M1}}_{ji}$ & A$^{{\rm M2}}_{ji}$  \\                     
\hline                                                                                                                                                                              
    5 &   28 &  2.675$+$01 &  0.000$+$00 &  0.000$+$00 &  0.000$+$00 &  2.239$+$03 &  0.000$+$00 &  0.000$+$00 \\                           
    5 &   30 &  1.724$+$01 &  0.000$+$00 &  0.000$+$00 &  0.000$+$00 &  1.019$+$04 &  0.000$+$00 &  0.000$+$00 \\                           
    5 &   31 &  1.591$+$01 &  0.000$+$00 &  0.000$+$00 &  0.000$+$00 &  9.928$+$02 &  3.595$+$02 &  0.000$+$00 \\                           
    6 &    7 &  2.427$+$02 &  0.000$+$00 &  0.000$+$00 &  0.000$+$00 &  5.483$-$01 &  0.000$+$00 &  0.000$+$00 \\                           
    6 &    9 &  2.064$+$02 &  0.000$+$00 &  0.000$+$00 &  0.000$+$00 &  1.590$+$01 &  2.528$+$02 &  0.000$+$00 \\                           
    6 &   10 &  1.525$+$02 &  9.856$+$05 &  1.717$-$06 &  3.448$-$06 &  0.000$+$00 &  0.000$+$00 &  4.256$-$02 \\                           
    6 &   11 &  1.503$+$02 &  0.000$+$00 &  0.000$+$00 &  0.000$+$00 &  2.866$+$02 &  2.797$+$06 &  0.000$+$00 \\
\hline                                                                                                                                                                              
\end{tabular}                                                                                                                                                                       
\end{table*}                                                                                                                                                                        
\newpage                                                                                                                                                                            
\clearpage                                                                                                                                                                          
\setcounter{table}{3}                                                                                                                                                               
\begin{table*}                                                                                                                                                                      
\caption{Transition wavelengths in vacuum ($\lambda_{ij}$ in $\rm \AA$), radiative rates (A$_{ji}$ in s$^{-1}$), oscillator strengths (f$_{ij}$, dimensionless), and line           
strengths (S, in atomic units), in Babushkin gauge, for electric dipole (E1), and A$_{ji}$ for E2, M1, and M2 transitions in W LVIII. ($a{\pm}b \equiv a{\times}$10$^{{\pm}b}$).}   
\begin{tabular}{rrrrrrrrrr}                                                                                                                                                         
\hline                                                                                                                                                                              
\hline                                                                                                                                                                              
$i$ & $j$ & $\lambda_{ij}$ & A$^{{\rm E1}}_{ji}$  & f$^{{\rm E1}}_{ij}$ & S$^{{\rm E1}}$ & A$^{{\rm E2}}_{ji}$  & A$^{{\rm M1}}_{ji}$ & A$^{{\rm M2}}_{ji}$  \\                     
\hline                                                                                                                                                                              
    7 &   27 &  3.359$+$01 &  0.000$+$00 &  0.000$+$00 &  0.000$+$00 &  7.847$+$05 &  4.118$+$07 &  0.000$+$00 \\                           
    7 &   28 &  3.156$+$01 &  0.000$+$00 &  0.000$+$00 &  0.000$+$00 &  4.742$+$04 &  0.000$+$00 &  0.000$+$00 \\                           
    7 &   30 &  1.911$+$01 &  0.000$+$00 &  0.000$+$00 &  0.000$+$00 &  1.735$+$02 &  0.000$+$00 &  0.000$+$00 \\                           
    7 &   31 &  1.750$+$01 &  0.000$+$00 &  0.000$+$00 &  0.000$+$00 &  5.989$+$03 &  4.714$+$03 &  0.000$+$00 \\                           
    8 &   11 &  4.253$+$02 &  0.000$+$00 &  0.000$+$00 &  0.000$+$00 &  3.254$+$01 &  0.000$+$00 &  0.000$+$00 \\                           
    8 &   13 &  3.103$+$02 &  0.000$+$00 &  0.000$+$00 &  0.000$+$00 &  2.268$+$01 &  0.000$+$00 &  0.000$+$00 \\                           
    8 &   17 &  4.559$+$01 &  0.000$+$00 &  0.000$+$00 &  0.000$+$00 &  1.992$+$03 &  0.000$+$00 &  0.000$+$00 \\
\hline                                                                                                                                                                              
\end{tabular}                                                                                                                                                                       
\end{table*}                                                                                                                                                                        
\newpage                                                                                                                                                                            
\clearpage                                                                                                                                                                          
\setcounter{table}{3}                                                                                                                                                               
\begin{table*}                                                                                                                                                                      
\caption{Transition wavelengths in vacuum ($\lambda_{ij}$ in $\rm \AA$), radiative rates (A$_{ji}$ in s$^{-1}$), oscillator strengths (f$_{ij}$, dimensionless), and line           
strengths (S, in atomic units), in Babushkin gauge, for electric dipole (E1), and A$_{ji}$ for E2, M1, and M2 transitions in W LVIII. ($a{\pm}b \equiv a{\times}$10$^{{\pm}b}$).}   
\begin{tabular}{rrrrrrrrrr}                                                                                                                                                         
\hline                                                                                                                                                                              
\hline                                                                                                                                                                              
$i$ & $j$ & $\lambda_{ij}$ & A$^{{\rm E1}}_{ji}$  & f$^{{\rm E1}}_{ij}$ & S$^{{\rm E1}}$ & A$^{{\rm E2}}_{ji}$  & A$^{{\rm M1}}_{ji}$ & A$^{{\rm M2}}_{ji}$  \\                     
\hline                                                                                                                                                                              
   10 &   19 &  4.471$+$01 &  0.000$+$00 &  0.000$+$00 &  0.000$+$00 &  0.000$+$00 &  0.000$+$00 &  5.128$+$01 \\                           
   10 &   20 &  4.432$+$01 &  3.917$+$08 &  2.307$-$04 &  6.731$-$05 &  0.000$+$00 &  0.000$+$00 &  5.554$+$01 \\                           
   10 &   21 &  4.319$+$01 &  3.001$+$09 &  8.396$-$04 &  2.388$-$04 &  0.000$+$00 &  0.000$+$00 &  0.000$+$00 \\                           
   10 &   23 &  3.883$+$01 &  2.681$+$09 &  1.212$-$03 &  3.099$-$04 &  0.000$+$00 &  0.000$+$00 &  4.059$+$04 \\                           
   10 &   24 &  3.833$+$01 &  0.000$+$00 &  0.000$+$00 &  0.000$+$00 &  0.000$+$00 &  0.000$+$00 &  3.545$+$04 \\                           
   10 &   26 &  3.752$+$01 &  0.000$+$00 &  0.000$+$00 &  0.000$+$00 &  0.000$+$00 &  0.000$+$00 &  1.025$+$04 \\                           
   10 &   28 &  3.419$+$01 &  7.999$+$11 &  2.803$-$01 &  6.310$-$02 &  0.000$+$00 &  0.000$+$00 &  8.496$+$02 \\
\hline                                                                                                                                                                              
\end{tabular}                                                                                                                                                                       
\end{table*}                                                                                                                                                                        
\newpage                                                                                                                                                                            
\clearpage                                                                                                                                                                          
\setcounter{table}{3}                                                                                                                                                               
\begin{table*}                                                                                                                                                                      
\caption{Transition wavelengths in vacuum ($\lambda_{ij}$ in $\rm \AA$), radiative rates (A$_{ji}$ in s$^{-1}$), oscillator strengths (f$_{ij}$, dimensionless), and line           
strengths (S, in atomic units), in Babushkin gauge, for electric dipole (E1), and A$_{ji}$ for E2, M1, and M2 transitions in W LVIII. ($a{\pm}b \equiv a{\times}$10$^{{\pm}b}$).}   
\begin{tabular}{rrrrrrrrrr}                                                                                                                                                         
\hline                                                                                                                                                                              
\hline                                                                                                                                                                              
$i$ & $j$ & $\lambda_{ij}$ & A$^{{\rm E1}}_{ji}$  & f$^{{\rm E1}}_{ij}$ & S$^{{\rm E1}}$ & A$^{{\rm E2}}_{ji}$  & A$^{{\rm M1}}_{ji}$ & A$^{{\rm M2}}_{ji}$  \\                     
\hline                                                                                                                                                                              
   12 &   23 &  3.979$+$01 &  0.000$+$00 &  0.000$+$00 &  0.000$+$00 &  2.021$+$05 &  8.147$+$07 &  0.000$+$00 \\                           
   12 &   24 &  3.927$+$01 &  0.000$+$00 &  0.000$+$00 &  0.000$+$00 &  1.071$+$06 &  1.968$+$06 &  0.000$+$00 \\                           
   12 &   26 &  3.842$+$01 &  0.000$+$00 &  0.000$+$00 &  0.000$+$00 &  6.677$+$03 &  1.084$+$08 &  0.000$+$00 \\                           
   12 &   27 &  3.744$+$01 &  0.000$+$00 &  0.000$+$00 &  0.000$+$00 &  1.447$+$06 &  0.000$+$00 &  0.000$+$00 \\                           
   12 &   28 &  3.493$+$01 &  0.000$+$00 &  0.000$+$00 &  0.000$+$00 &  3.443$+$06 &  4.687$+$07 &  0.000$+$00 \\                           
   12 &   29 &  3.399$+$01 &  0.000$+$00 &  0.000$+$00 &  0.000$+$00 &  1.502$+$06 &  1.413$+$08 &  0.000$+$00 \\                           
   12 &   30 &  2.030$+$01 &  0.000$+$00 &  0.000$+$00 &  0.000$+$00 &  1.642$+$03 &  5.083$+$03 &  0.000$+$00 \\
\hline                                                                                                                                                                              
\end{tabular}                                                                                                                                                                       
\end{table*}                                                                                                                                                                        
\newpage                                                                                                                                                                            
\clearpage                                                                                                                                                                          
\setcounter{table}{3}                                                                                                                                                               
\begin{table*}                                                                                                                                                                      
\caption{Transition wavelengths in vacuum ($\lambda_{ij}$ in $\rm \AA$), radiative rates (A$_{ji}$ in s$^{-1}$), oscillator strengths (f$_{ij}$, dimensionless), and line           
strengths (S, in atomic units), in Babushkin gauge, for electric dipole (E1), and A$_{ji}$ for E2, M1, and M2 transitions in W LVIII. ($a{\pm}b \equiv a{\times}$10$^{{\pm}b}$).}   
\begin{tabular}{rrrrrrrrrr}                                                                                                                                                         
\hline                                                                                                                                                                              
\hline                                                                                                                                                                              
$i$ & $j$ & $\lambda_{ij}$ & A$^{{\rm E1}}_{ji}$  & f$^{{\rm E1}}_{ij}$ & S$^{{\rm E1}}$ & A$^{{\rm E2}}_{ji}$  & A$^{{\rm M1}}_{ji}$ & A$^{{\rm M2}}_{ji}$  \\                     
\hline                                                                                                                                                                              
   15 &   17 &  6.585$+$02 &  0.000$+$00 &  0.000$+$00 &  0.000$+$00 &  2.886$-$01 &  0.000$+$00 &  0.000$+$00 \\                           
   15 &   19 &  2.378$+$02 &  0.000$+$00 &  0.000$+$00 &  0.000$+$00 &  5.438$+$00 &  0.000$+$00 &  0.000$+$00 \\                           
   15 &   20 &  2.270$+$02 &  0.000$+$00 &  0.000$+$00 &  0.000$+$00 &  2.607$+$01 &  6.652$+$03 &  0.000$+$00 \\                           
   15 &   21 &  2.003$+$02 &  0.000$+$00 &  0.000$+$00 &  0.000$+$00 &  0.000$+$00 &  1.719$+$04 &  0.000$+$00 \\                           
   15 &   23 &  1.317$+$02 &  0.000$+$00 &  0.000$+$00 &  0.000$+$00 &  3.039$+$02 &  1.276$+$06 &  0.000$+$00 \\                           
   15 &   24 &  1.261$+$02 &  0.000$+$00 &  0.000$+$00 &  0.000$+$00 &  4.039$+$02 &  0.000$+$00 &  0.000$+$00 \\                           
   15 &   26 &  1.178$+$02 &  0.000$+$00 &  0.000$+$00 &  0.000$+$00 &  3.931$-$01 &  0.000$+$00 &  0.000$+$00 \\
\hline                                                                                                                                                                              
\end{tabular}                                                                                                                                                                       
\end{table*}                                                                                                                                                                        
\newpage                                                                                                                                                                            
\clearpage                                                                                                                                                                          
\setcounter{table}{3}                                                                                                                                                               
\begin{table*}                                                                                                                                                                      
\caption{Transition wavelengths in vacuum ($\lambda_{ij}$ in $\rm \AA$), radiative rates (A$_{ji}$ in s$^{-1}$), oscillator strengths (f$_{ij}$, dimensionless), and line           
strengths (S, in atomic units), in Babushkin gauge, for electric dipole (E1), and A$_{ji}$ for E2, M1, and M2 transitions in W LVIII. ($a{\pm}b \equiv a{\times}$10$^{{\pm}b}$).}   
\begin{tabular}{rrrrrrrrrr}                                                                                                                                                         
\hline                                                                                                                                                                              
\hline                                                                                                                                                                              
$i$ & $j$ & $\lambda_{ij}$ & A$^{{\rm E1}}_{ji}$  & f$^{{\rm E1}}_{ij}$ & S$^{{\rm E1}}$ & A$^{{\rm E2}}_{ji}$  & A$^{{\rm M1}}_{ji}$ & A$^{{\rm M2}}_{ji}$  \\                     
\hline                                                                                                                                                                              
   18 &   20 &  3.603$+$02 &  0.000$+$00 &  0.000$+$00 &  0.000$+$00 &  4.496$+$00 &  0.000$+$00 &  0.000$+$00 \\                           
   18 &   22 &  1.922$+$02 &  0.000$+$00 &  0.000$+$00 &  0.000$+$00 &  1.028$+$01 &  2.661$+$04 &  0.000$+$00 \\                           
   18 &   23 &  1.676$+$02 &  0.000$+$00 &  0.000$+$00 &  0.000$+$00 &  5.703$+$00 &  0.000$+$00 &  0.000$+$00 \\                           
   18 &   24 &  1.588$+$02 &  0.000$+$00 &  0.000$+$00 &  0.000$+$00 &  2.588$+$01 &  4.310$+$04 &  0.000$+$00 \\                           
   18 &   25 &  1.594$+$02 &  0.000$+$00 &  0.000$+$00 &  0.000$+$00 &  4.167$+$02 &  2.612$+$06 &  0.000$+$00 \\                           
   18 &   26 &  1.458$+$02 &  0.000$+$00 &  0.000$+$00 &  0.000$+$00 &  1.497$+$02 &  9.400$+$04 &  0.000$+$00 \\                           
   18 &   27 &  1.326$+$02 &  0.000$+$00 &  0.000$+$00 &  0.000$+$00 &  6.433$+$02 &  8.938$+$05 &  0.000$+$00 \\
\hline                                                                                                                                                                              
\end{tabular}                                                                                                                                                                       
\end{table*}                                                                                                                                                                        
\newpage                                                                                                                                                                            
\clearpage                                                                                                                                                                          
\setcounter{table}{3}                                                                                                                                                               
\begin{table*}                                                                                                                                                                      
\caption{Transition wavelengths in vacuum ($\lambda_{ij}$ in $\rm \AA$), radiative rates (A$_{ji}$ in s$^{-1}$), oscillator strengths (f$_{ij}$, dimensionless), and line           
strengths (S, in atomic units), in Babushkin gauge, for electric dipole (E1), and A$_{ji}$ for E2, M1, and M2 transitions in W LVIII. ($a{\pm}b \equiv a{\times}$10$^{{\pm}b}$).}   
\begin{tabular}{rrrrrrrrrr}                                                                                                                                                         
\hline                                                                                                                                                                              
\hline                                                                                                                                                                              
$i$ & $j$ & $\lambda_{ij}$ & A$^{{\rm E1}}_{ji}$  & f$^{{\rm E1}}_{ij}$ & S$^{{\rm E1}}$ & A$^{{\rm E2}}_{ji}$  & A$^{{\rm M1}}_{ji}$ & A$^{{\rm M2}}_{ji}$  \\                     
\hline                                                                                                                                                                              
   22 &   24 &  9.128$+$02 &  0.000$+$00 &  0.000$+$00 &  0.000$+$00 &  1.272$-$01 &  3.567$+$02 &  0.000$+$00 \\                           
   22 &   25 &  9.345$+$02 &  0.000$+$00 &  0.000$+$00 &  0.000$+$00 &  8.593$-$04 &  5.308$+$03 &  0.000$+$00 \\                           
   22 &   26 &  6.029$+$02 &  0.000$+$00 &  0.000$+$00 &  0.000$+$00 &  4.779$+$00 &  2.766$+$03 &  0.000$+$00 \\                           
   22 &   27 &  4.269$+$02 &  0.000$+$00 &  0.000$+$00 &  0.000$+$00 &  3.648$+$00 &  1.047$+$03 &  0.000$+$00 \\                           
   22 &   28 &  2.348$+$02 &  0.000$+$00 &  0.000$+$00 &  0.000$+$00 &  7.962$-$06 &  0.000$+$00 &  0.000$+$00 \\                           
   22 &   30 &  4.016$+$01 &  0.000$+$00 &  0.000$+$00 &  0.000$+$00 &  9.913$+$02 &  0.000$+$00 &  0.000$+$00 \\                           
   22 &   31 &  3.364$+$01 &  0.000$+$00 &  0.000$+$00 &  0.000$+$00 &  8.414$+$05 &  3.364$+$08 &  0.000$+$00 \\
\hline                                                                                                                                                                         
\end{tabular}                                                                                                                                                                  
\end{table*}                                                                                                                                                                   
 

\newpage
\clearpage

\begin{flushleft}
Table 5. Lifetimes ($\tau$, s) for the levels of W LVIII and the {\em dominant} A- values. $a{\pm}b \equiv a{\times}$10$^{{\pm}b}$.
\end{flushleft}
\begin{tabular}{rllrrrlrrrrrr} \hline
 & & & & & & & &  \\
Index  & Configuration             & Level                &  GRASP1a &  GRASP1b & GRASP2    & Dominant A- values (s$^{-1}$) \\
& & & & & & & &   \\ \hline
& & & & & & & &   \\
  1  &  3s$^2$3p$^5$               &  $^2$P$^o_{3/2}$	  & ......    & ......    & ......     & ........			      \\
  2  &  3s$^2$3p$^4$($^3$P)3d 	   &  $^4$D$  _{3/2}$	  & 3.73-10   & 3.737-10  & 3.779-10   & 1--2 E1=2.68+09		      \\
  3  &  3s$^2$3p$^4$($^3$P)3d      &  $^4$D$  _{5/2}$	  & 2.40-10   & 2.399-10  & 2.406-10   & 1--3 E1=4.17+09		      \\
  4  &  3s$^2$3p$^4$($^3$P)3d      &  $^4$P$  _{1/2}$	  & 4.89-11   & 4.898-11  & 4.972-11   & 1--4 E1=2.04+10		      \\
\hline            								                	 
\end{tabular}   								   					       
			      							   					       
\begin{flushleft}													       
{\small
GRASP1a: earlier calculations of Mohan et al. \cite{mas}  from the {\sc grasp} code with 1163 levels \\										
GRASP1b: present calculations  from the {\sc grasp} code with 1163 levels \\											       
GRASP2: present calculations from the {\sc grasp} code with 3749 levels\\															       
}															       
\end{flushleft} 
						
\end{document}